\documentclass{edp-jp4}
\usepackage{graphicx}
\begin{document}
\title{GAIA Spectroscopy and Radial Velocities} 
\author{Ulisse Munari}\address{Osservatorio Astronomico di Padova, 
Sede di Asiago, I-36012 Asiago (VI), Italy}
\maketitle
\begin{abstract} 
GAIA spectroscopic and radial velocity performancies are reviewed on the
base of ground-based test observations and simulations. The prospects for
accurate analysis of stellar atmospheres (temperature, gravity, chemical
abundances, rotation, peculiarities) and precise radial velocities (single
stars, binaries, pulsating stars) are colorful provided the spectral
dispersion is high enough. A higher dispersions also favors a given
precision of radial velocities to be reached at fainter magnitudes: for
example, with current parameters for GAIA spectrograph, a 1 km~sec$^{-1}$
accuracy on epoch RVs of a K0 star is reached at $V\sim$13.0 mag with
0.25~\AA/pix dispersion spectra, at $V\sim$10.3 mag for 0.5~\AA/pix, and $V\sim$6.7 mag
for 1~\AA/pix. GAIA radial velocities for single stars can match
the $\sim$0.5 km~sec$^{-1}$ mean accuracy of tangential motions at $V=15$
mag, provided the observations are performed at a dispersion not less than
0.5 \AA/pix.
\end{abstract}

\section{Introduction}

The giant leap that GAIA spectroscopy will lead us through can be
sized by four basic considerations: ($a$) GAIA will record multi-epoch
spectra for a magnitude complete sample of stars $\sim 10^3$ larger than any
whole-sky existing database (e.g. HD survey, progressing Michigan project,
etc.); ($b$) for each target, an average of 67 epoch spectra will be
recorded over the five year mission lifetime; ($c$) the wavelength and flux
calibrated spectra will be available in digital format to the community;
($d$) the foreseeable spectral dispersion (0.75 \AA/pix are currently
baselined) is significantly higher that those of other whole-sky surveys.

A review of GAIA spectroscopy has already been presented by Munari (1999a,
hereafter M99a). We will consider here mainly updates to the content of M99a
reflecting advancements in some areas over the last couple of years.
Therefore, to cope with the generous but limited amount of space available
to this review in its printed format, basic physics and overall
considerations developed in M99a will not be discussed here again. Technical
aspects connected to spacecraft optical and mechanical assembly, telemetry
budgets, modus operandi, limiting magnitudes etc., are covered in the ESA's
{\sl GAIA Concept and Technology Study Report} (ESA-SCI-2000-4) and in
abridged format in Perryman et al. (2001).

GAIA spectra will cover the 8490-8750 \AA\ wavelength range, centered on the
near-IR Ca~II triplet and head of the Paschen series. The extention to 8750
\AA\ (the redder Ca~II line laying at 8662.141 \AA) allows observation of
remarkable N~I \#1 and 8 multiplets in hot stars and particularly strong
Fe~I, Mg~I and Ti~I lines in cool stars. GAIA wavelength range is the only
spectral window in the near-IR which is not crunched by telluric absorptions (cf.
Figure~1 of M99a), allowing uncontaminated ground-based preparatory and
follow-up observations.

Ca~II triplet is by far the strongest line feature in the red/near-IR
spectra of cool stars (cf. Fig~1 of Munari \& Castelli 2000; Jaschek \&
Jaschek 1995), being core-saturated even in metal-poor halo stars, thus
allowing derivation of radial velocities on epoch-spectra of even the
faintest and more metal deprived GAIA spectral targets. Cool stars will
vastly dominate among the latter (with O and B stars barely traceable). 
GAIA wavelength range (restricted to $\bigtriangleup \lambda \sim 250$ \AA\
by optical and telemetry constraints) is placed toward peak emission in cool
stars and lower interstellar extinction, with obvious benefits for the
number of detectable targets.

The number of GAIA spectral targets (approaching 10$^8$ in current
estimates), will require fully automatic data treatment and analysis. Line
intensities and ratios may be useful in pre-flight ground-based preparatory
work and/or quick-look classification and re-direction along the reduction
pipeline of actual GAIA data. However, it is clear that proper exploitation
of GAIA spectra will required a smart spectral synthesis approach. Even if
currently available synthetic models of stellar atmosphere (MARCS, ATLAS,
Phoenix, etc.) and nebular regions (CLOUDY, etc.) will be probably quite
improved by the time GAIA data will be available (and new families of models
will probably be developed too), nevertheless they play a fundamental role
right now in the current infancy of GAIA spectroscopy, by offering precious
guidelines, ways to improve basic underlying physics (for ex. atomic
constants) and unlimited databases for simulations.

Most of GAIA performances will depend on the eventually  adopted spectral
dispersion. An example of how lowering the resolution affects spectral
appearance of a K0~III star -- a typical GAIA target -- is given in
Figure~1. On one side the race for fainter limiting magnitudes and smallest
demand on telemetry push for mid to low spectral dispersions. On the other
side, getting the best astrophysical return and the highest accuracy of
radial velocities most decidedly ask for high dispersions.  The best
compromise will have to balance between them. 

\begin{figure}[b]
\includegraphics[width=14.5cm]{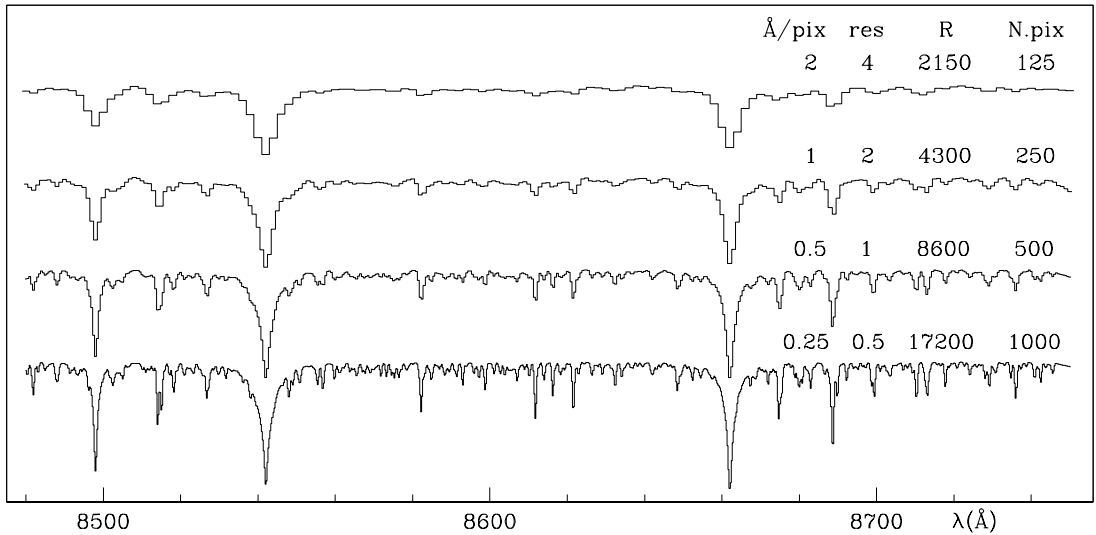}
\caption{The same K0~III spectrum at different dispersions/resolutions. The
number in the four columns give the dispersion (in \AA/pix), the resolution (in
\AA, with FWHM(PSF)=2 pix), the resolving power and the number of pixel
required to cover the whole GAIA wavelength range.}
\end{figure}

\begin{figure}[b]
\includegraphics[width=14.5cm]{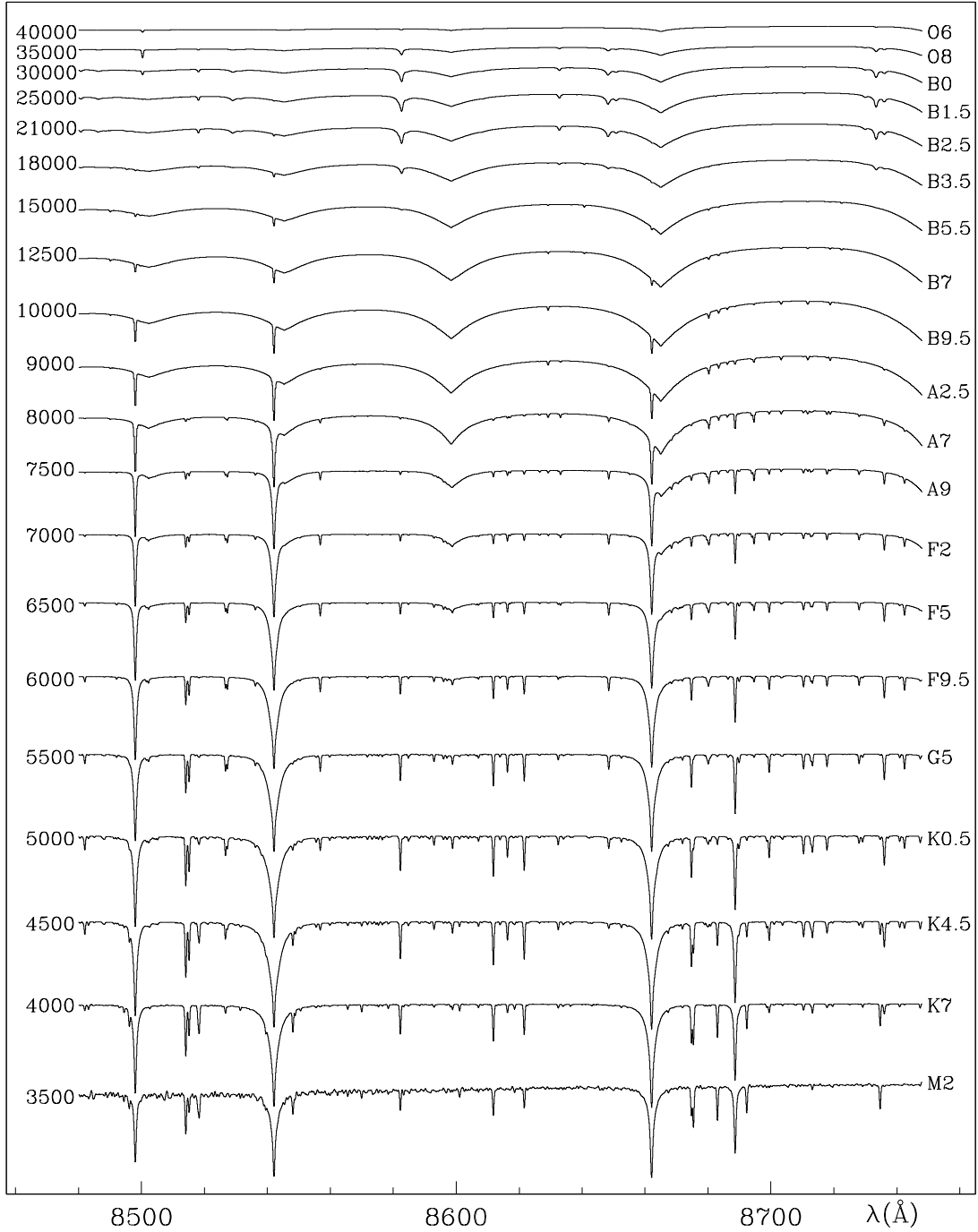}
\caption{Sequence of synthetic spectra (from Munari \& Castelli 2000,
Castelli \& Munari 2001) illustrating the variations along the main
sequence ($T_{eff}$ in K on the left and corresponding spectral type for
luminosity class V on the right) for moderately metal poor stars
([Z/Z$_\odot$]=$-$0.5). All spectra are on the same ordinate scale,
only displaced in their zero-points.}
\end{figure}

\clearpage

The spectra presented in this review carry a 0.25~\AA/pix dispersion, at the
high end of the 0.25$-$1.0~\AA/pix range currently considered, thus allowing
the reader to guess the highest possible GAIA performancies. In the
following, effective temperatures and surface gravities for MKK spectral
types are adopted from Strayzis (1992).

\section{Stellar physical and classification parameters}

\subsection{Temperature/spectral type}

A temperature sequence spanning the MKK classification scheme is presented
in Figure~2. M to F spectral types are governed by the Ca~II triplet, hotter
ones by He~I, N~I and the hydrogen Paschen series. A rich forest of metal
lines populates the GAIA wavelength range (cf. Figure~3 in M99a), which is
dominated by Ca~II, Fe~I, Ti~I atomic lines and CN molecular transitions. Relevant
absorptions are also due to Mg~I, Si~I, Cr~I, N~I, Co~I, Ni~I, Mn~I, S~I as
well as TiO, with other elements and molecules contributing weaker spectral
signatures.

Such a harvest makes spectral classification over the $\bigtriangleup\lambda
\sim$250~\AA\ GAIA range nearly as much easy as it is for the
$\bigtriangleup\lambda \sim$1000~\AA\ classical MKK range (which extends from
3900 to 4900~\AA). Only O and B stars perform less good, which is
however of no concern given their barely traceable fraction among GAIA targets.

Diagnostic line ratios useful for spectral classification purposes can be
easily derived on GAIA spectra. Two examples of line ratios are illustrated in Figure~3.
Near-IR Ca~II over Paschen lines are highly effective in classifying late B, A and F
stars, as it if for Ca~II H and K over Balmer lines in optical spectra.
Ti 8674.7/FeI 8675.4 works very well in G, K and M stars, with the
following expression giving the fitting curve in Figure~3 (right panel):
\begin{equation}
R\ =\frac{\rm Ti\,I\ 8674.7\ \AA}{\rm Fe\,I\ 8675.4\ \AA}\ \Longrightarrow\ \ \ \ 
T_{eff}\ =\ 6072\  -\ 2188\times R\ +\  364\times R^2
\end{equation}
A $\bigtriangleup R =$4\% error in the line ratio (typical for the
observations in Figure~3) corresponds to just $\bigtriangleup T_{eff} =$ 65
K. It worth to mention that also GAIA photometry will (obviously) estimate
the temperature of target stars (however with contamination from
interstellar reddening, if present), providing independent data to be
compared with spectroscopic findings.
  
\subsection{Gravity/luminosity class}

Figure~5 illustrates line behavior with luminosity class. With lowering
surface gravity (increasing luminosity, decreasing pressure) the intensity
of absorption lines goes up, as much as in classical optical spectra and
required by physics. The equivalent width of Ca~II lines shows a pronounced
positive luminosity effect (collisional de-population of excited state less
effective with decreasing pressure). Width of Paschen lines presents a
negative luminosity effect (being pressure broadened, similarly to Balmer
lines) as given in Figure~4. It is again worth noticing that surface
gravities will be also quite effectively measured by GAIA combining
astrometric distances with photometry (cf. sect. 2.1 in Munari 1999b,
hereafter M99b).

As for temperature/spectral type, it is easy to derive diagnostic line ratios
highly sensitive to surface gravity/luminosity class. The steep behavior of
(Si~I 8728.0 + Fe~I 8729.1)/(Mg~I 8736.0) in G5 stars is illustrated in
Figure~4, with fitting curve given by:

\begin{equation}
R\ =\ \frac{\rm Si\,I\ +\ Fe\,I\ 8728\ \AA}{\rm Mg\,I\ 8736\ \AA}\ \Longrightarrow\ \ \ \ \ 
\lg\, g\ =\ 3.6\ +\ 3.0\times R\
-\ 2.57\times R^2
\end{equation}
A $\bigtriangleup R =$3\%  error (typical for Figure~4 data)
corresponds to just $\bigtriangleup \lg g$=0.11

\clearpage

\begin{figure}[b]
\includegraphics[width=14.5cm]{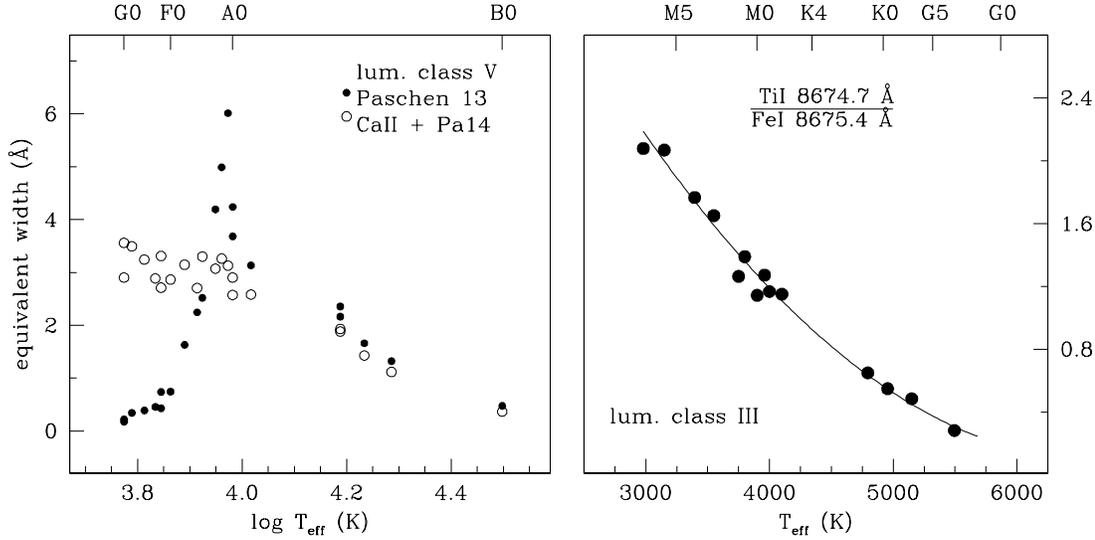}
\caption{Example of temperature diagnostic ratios (Boschi 2000, Boschi \&
Munari 2001) from real spectra (from Munari \& Tomasella 1999). {\sl Left
panel:} hydrogen Paschen and Ca~II lines in main sequence stars. {\sl Right
panel :} a powerful temperature diagnostic ratio for cool giants (requiring
0.25 \AA/pix dispersion spectra to resolve the lines)}
\end{figure}

\begin{figure}[b]
\includegraphics[width=14.5cm]{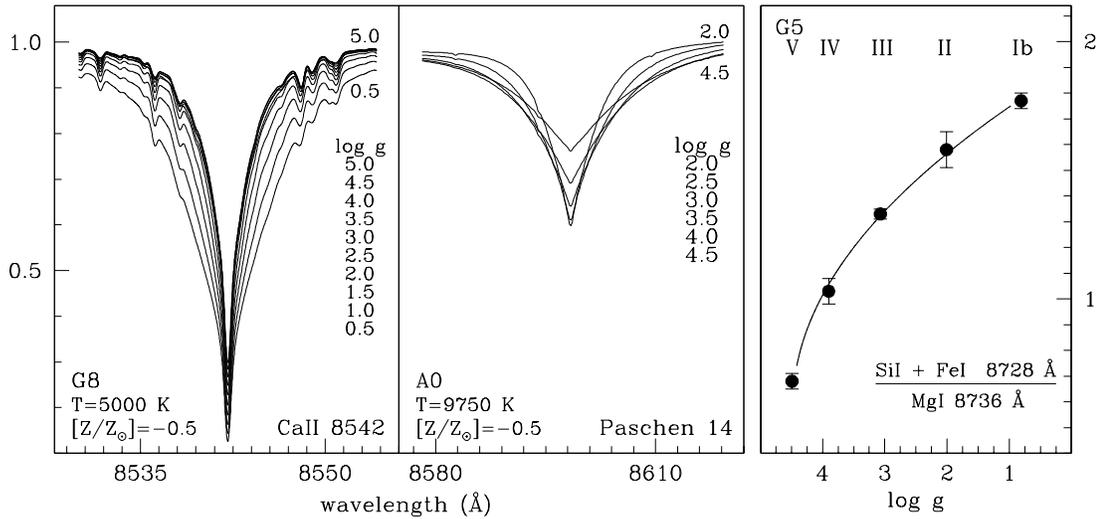}
\caption{{\sl Left:} gravity effects on the profile of CaII~8542 \AA\
for $T_{eff}$=5000~K ($\sim$G8 spectral type) and
[Z/Z$_\odot$]=$-$0.5 (synthetic spectra from Munari \& Castelli 2000). 
{\sl Center:} gravity effects on the profile of Paschen 14, 8598 \AA\
for $T_{eff}$=9750~K ($\sim$A0 spectral type) and
[Z/Z$_\odot$]=$-$0.5 (synthetic spectra from Zwitter et al. 2001). 
{\sl Right:} dependence upon gravity of the ratio (SiI~8728.0 + FeI~8729.1)/MgI~8736.0 
(Boschi and Munari 2001) on G5 real spectra (from Munari \& Tomasella 1999).}
\end{figure}

\clearpage

\begin{figure}[b]
\includegraphics[width=14.5cm]{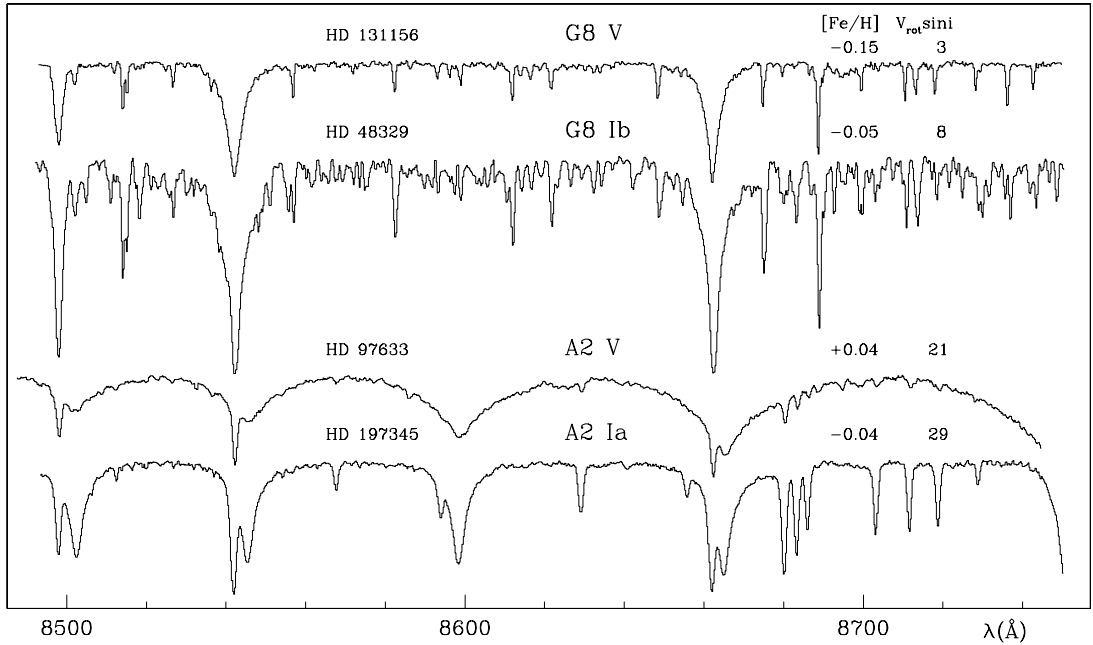}
\caption{{\sl Left:} gravity effects on G8 and A2 spectra (from Munari \&
Tomasella 1999).} 
\end{figure}

\begin{figure}[b]
\includegraphics[width=14.5cm]{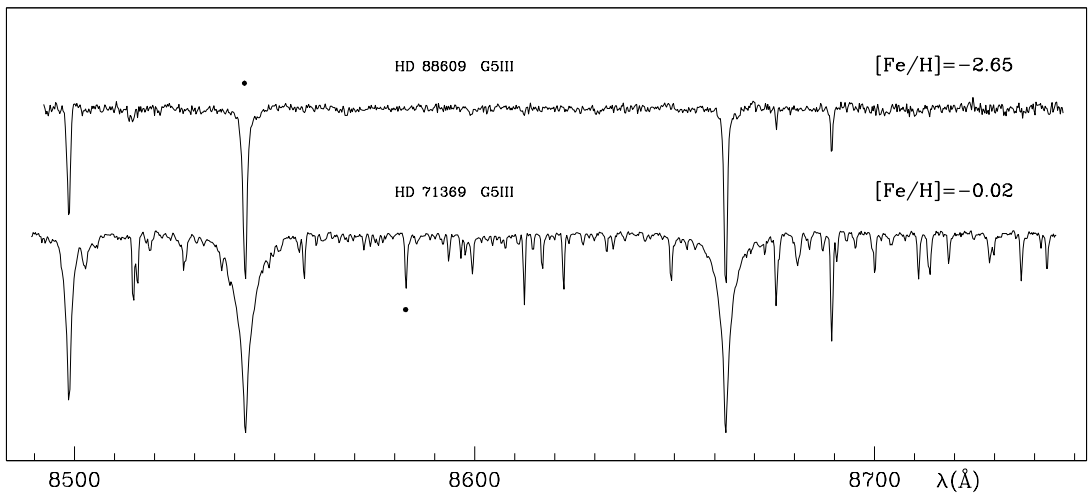}
\caption{Examples of G5~III stars of quite different metallicities (from
Marrese 2000). On cool stars, even at very low abundances,  Ca~II lines
remain strong, allowing accurate radial velocity measurements, while nearly
all other metallic lines have gone.}
\end{figure}

\clearpage

\subsection{Chemical abundances}

The intensity of absorption lines obviously correlates with the chemical
abundances. Figure~6 presents spectra of two G5~III stars of widely differing
metallicities ([Fe/H]=$-$2.65 and $-$0.02). 

Even at the lowest metallicities (like those found in the Halo and globular
clusters), Ca~II lines remain core-saturated while nearly all other metallic
lines have gone, allowing accurate radial velocity measurements and fine
rotational velocity estimates.

The possibility to perform chemical analysis on the recorded spectra will
depend on the spectral dispersion that will be eventually adopted for GAIA.
Figure~1 is illuminating in this sense. At 0.25 \AA/pix hundreds of
absorptions lines blossom over the whole GAIA range, tracing tens of
different elements which {\sl individual} chemical abundances can be derived
as routinely done with high resolution optical spectra secured with
ground-based telescopes. At 0.5 \AA/pix, lines from different elements
merge into blends and individual chemical abundances can be derived (with a
limited accuracy) probably only for Fe~I, Ti~I and Mg~I (tests are
underway).  At even lower dispersions, chemical analysis looks hopeless,
with absorption lines washing into an unfeatured continuum.

It is also worth to mention that GAIA narrow band photometry (M99b;
V.Vansevicius, M.Grenon, these proceedings) should estimate metallicity (a
weighted average of individual chemical abundances) from color indexes with
a sensivity comparable to best performing, existing ground-based photometric
systems (Moro \& Munari 2000).

\section{Rotational velocities}

Projected rotational velocity ($V_{rot} \sin i$) is another intrinsic
property of stars that GAIA can measure with relevant confidence, provided the
spectral dispersion is high enough. The impact of axial rotation on stellar
evolution is only recently being appreciated and modeled (cf. A.Maeder,
this volume; Maeder \& Meynet 2001 and references therein).

\begin{figure}[Hb!]
\includegraphics[width=14.5cm]{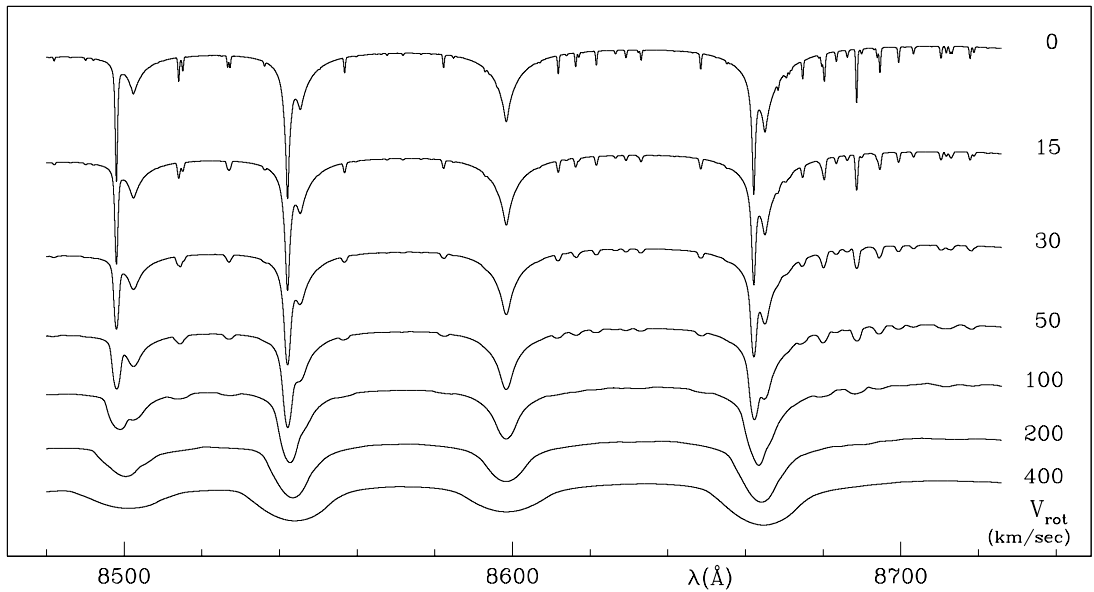}
\caption{Rotational velocity sequence for F0~III giants (spectra from Zwitter 
et al. 2001).}
\end{figure}

A sequence of rotationally broadened F0~III spectra is presented in
Figure~7. Ca~II and Paschen lines perform equally well up to very high
$V_{rot}$, offering bright prospects for measuring rotational velocities
both in cool and hot stars, respectively. Other, weaker lines are useless at
$V_{rot} \geq$40~km~sec$^{-1}$, being washed out in the adjacent
continuum.

Rotational velocities of G-K-M stars (constituting the largest fraction of
GAIA targets) are usually confined to $V_{rot} \leq$15~km~sec$^{-1}$
(Glebocki et al. 2000). To effectively discriminate at so low values, a
properly high spectral resolution is necessary. 0.5 \AA/pix spectra
would not allow GAIA to distinguish $V_{rot}\sin i =$15~km~sec$^{-1}$ stars from
non-rotating ones, i.e. rotational velocities will be undetermined for the
majority of GAIA targets. Instead, 0.25 \AA/pix spectra can detect and
measure rotational velocities to an accuracy of $\bigtriangleup V_{rot}\sin
i =$5~km~sec$^{-1}$, thus providing decent sensitivity to rotation
for {\sl all} GAIA targets.  The rotational broadening of narrow lines 
may be expressed in term of the total width at half
intensity (in \AA) as
\begin{equation}
V_{rot}\sin i \ =\ 42.4 \times HIW \ -\ 35 \ \ \ \ \ \ \ \  km \ sec^{-1}
\end{equation}

\section{Radial velocities}

\subsection{Single stars}

Munari et al. (2001a, hereafter M01a) have investigated in detail GAIA
radial velocity performances as function of spectral resolution and
signal-to-noise ratio, by obtaining 782 real spectra and using them as
inputs for 6700 automatic cross-correlation runs. M01a have explored the
dispersions 0.25, 0.5, 1 and 2 \AA/pix (bracketing the 0.75 \AA/pix
currently baselined) over S/N ranging from 12 to 110,  carefully
maintaining the condition FWHM (PSF) = 2 pixels. M01a
have investigated late-F to early-M stars (constituting the vast majority of
GAIA targets), slowly rotating ($<\, V_{rot}\, \sin\, i\,>$ = 4
km~sec$^{-1}$, as for field stars at these spectral types), of solar
metallicity ($<$\,[Fe/H]\,$>$ = $-$0.07) and not binary (target stars
selected among IAU standard radial velocity stars). The results are
accurately described by:
\begin{equation}
\lg\, \sigma\, =\, 0.6\times(\lg\frac{S}{N})^2\, -\, 2.4\times\lg\frac{S}{N}\, 
+\, 1.75\times \lg D\, +\, 3
\end{equation}
where $\sigma$ is the cross-correlation standard error (in km sec$^{-1}$)
and $D$ is the spectral dispersion (in \AA/pix). Inserting 0.25, 0.5, 1, 2
for $D$ in the above expression gives the four fitting curves in Figure~8. The
spectral dispersion is the dominant factor governing the
accuracy of radial velocities, with S/N being less important. 
Spectral mis-match affects the results only at the highest S/N. M01a findings
suggests that mission-averaged GAIA radial velocities on non-variable, single
stars can match the $\sim$0.5 km~sec$^{-1}$ mean accuracy of
tangential motions at $V=15$ mag, provided the observations are performed at
a dispersion not less than 0.5 \AA/pix. Binary and/or fast rotating and/or
pulsating and/or surface spotted stars will require higher accuracies
(thus higher spectral dispersions) in order to disentangle perturbing
effects from baricentric motion.

The flattening of RV performances at the highest S/N in Figure~8 and
Eq.(4.1) results from the spectral type mismatch between template and
program stars in M01a study (while metallicities and rotational velocities
were instead pretty similar and therefore not influent).  The large but
however limited number of real spectra (782) and the extreme paucity of IAU
RV standards sharing the same spectral classification, forced spectral type
mismatch in M01a investigation. Extensive simulations on huge numbers of
synthetic spectra have been performed by Zwitter (2001) to get rid of
mismatches unavoidable with real spectra. His results confirm Eq. (4.1) and
Figure~8 behavior at low and mid S/N. As expected, they do not show the
upward curvature at the highest S/N induced in Figure~8 by the spectral
mismatch (see also D. Katz, these proceedings, for additional results on
GAIA radial velocities).

M01a and Zwitter (2001) investigations argue in favor of increasing the
spectral dispersion to get more accurate radial velocities, for a fixed
photon budget (like in GAIA fixed exposure time observations). From the
results with real spectra of Figure~8, with current parameters for GAIA
spectrograph, a 1 km~sec$^{-1}$ accuracy on epoch RVs of a K0 star is
reached at $V\sim$13.0 mag with 0.25~\AA/pix spectra, at $V\sim$10.3 mag for
0.5~\AA/pix, and $V\sim$6.7 mag for 1~\AA/pix dispersion. The reason for
this is quite obvious looking at Figure~1: at high dispersions {\sl all}
pixels carry RV information, even at low S/N, while at low dispersion only
{\sl a few} pixels carry RV information, no matter how high is the S/N.

\begin{figure}[Ht!]
\parbox{9.5cm}{\includegraphics[width=9.0cm]{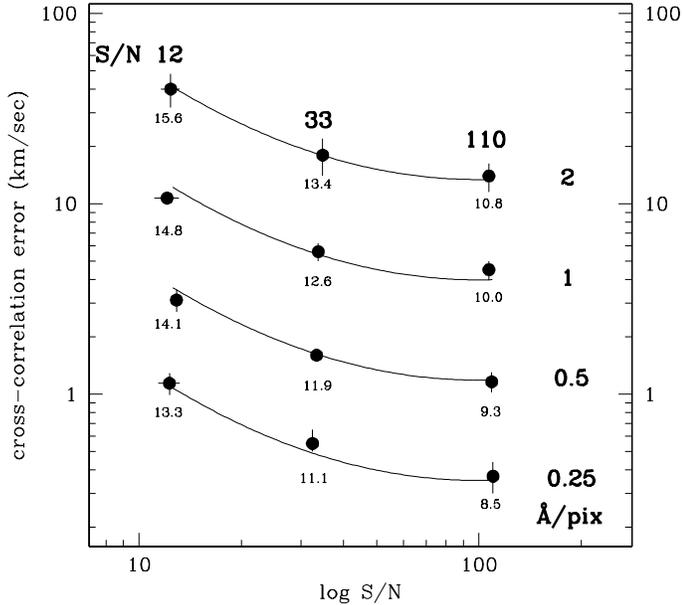}}
\parbox{5.0cm}{\caption{The accuracy of radial velocities obtained via
cross-correlation for late-F to early-M stars as function of S/N and
spectral dispersion (Agnolin 2000; Munari et al. 2001a). 
The results based on 782 real spectra and 6700 cross-correlation runs 
include the effect of mis-match between object and
template stars (removing it would reduce the upward curvature at higher S/N).
The $V$ magnitude of unreddened G5~V stars producing epoch spectra of the
given S/N is reported next to the points (computed for: 
mirror size = 75$\times$70 cm; overall
throughput = 35\%; crossing time = 60.8 sec; $I_{\rm C}^{mag=0.0}$ =
1.196$\times 10^{-9}$ erg cm$^{-2}$ sec$^{-1}$ \AA$^{-1}$ = 520 photons 
cm$^{-2}$ sec$^{-1}$ \AA$^{-1}$ ; R.O.N. = 3
$e^{-1}$; dark = 0.01 $e^{-1}$ sec$^{-1}$; background $I_{\rm C}$=21.5
arcsec$^{-2}$).}}
\end{figure}

\begin{figure}[Hb!]
\includegraphics[width=14.5cm]{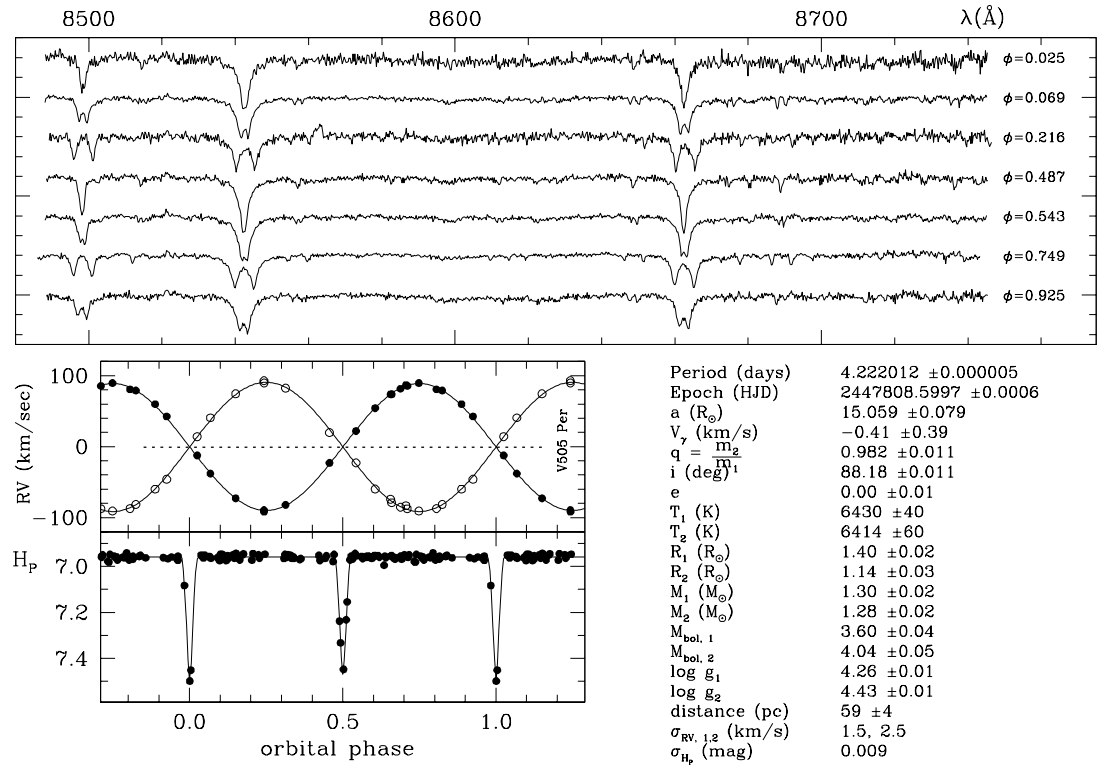}
\caption{V505 Per as an example of GAIA performance on eclipsing binaries
(from Munari et al. 2001b). {\sl Upper panel:} a sample of the recorded spectra to
illustrate line splitting with orbital phase. {\sl Lower left panel:} radial
velocity curve and model fitting, Hipparcos $H_P$ photometry and model
fitting. {\sl Lower right panel:} orbital and physical parameters as derived
from modeling of radial velocity and photometric data.}
\end{figure}

\subsection{Binary stars}

Eclipsing stars represent the most astrophysically relevant type of
binary stars GAIA will spectroscopically observe during its mission.

They are a prime tool to derive fundamental stellar parameters like mass and
radius, or the temperature scale. Moreover, their use as accurate, geometric
distance indicators is rapidly growing (current best distances to LMC, dwarf
spheroidals, globular clusters). Their study is by no means a simple task as
evidenced by the fact that stellar parameters have been derived with an
accuracy of 1\% or better for less than a hundred objects. Scaling the
Hipparcos results (0.8\% of the 118218 stars surveyed turned out to be
eclipsing), $\sim4\cdot 10^5$ of all $V\leq 15$ mag GAIA targets should be
eclipsing binaries, with a average G7 spectral type. It may be estimated
that about 25\% of the eclipsing binaries will be
double-lined in GAIA spectral observations (cf. Carquillat et al. 1982),
thus $\sim1\cdot 10^5$ of all $V\leq 15$ mag GAIA targets.  Even if for only
5\% of them it should be possible to derive orbits and stellar parameters at
1\% precision, this still would be $100\times$ what
humanity have so far collected from devoted ground-based efforts during the
whole last century (cf. Andersen 1991).

That a 1\% accuracy in orbital and physical parameters of eclipsing binaries
is feasible within GAIA has been demonstrated by Munari et al. (2001b,
hereafter M01b). 0.25 \AA/pix spectra over the GAIA range have been secured
from the ground for 15 eclipsing binaries (mostly unstudied in literature
and distributed among detached, semidetached, contact and intrinsic variable
types) and combined with Hipparcos $V_T$, $B_T$, $H_P$ photometry to
properly simulate GAIA data harvest.

Example results from M01b for V505~Per are given in Figure~9. Semi-major
axis, masses, surface gravities, effective temperatures, inclination,
eccentricity, baricentric velocity, epoch and orbital period are in the
1\% or better accuracy regime. Radii of individual components (and therefore
individual bolometric magnitudes) however depends on how well the branches
of eclipses are mapped. In the case of Hipparcos observations of V505~Per
{\sl only three points} cover the principal eclipse. Even if model solution
can fit these three points to a formal accuracy of $\sim$1\% (cf.
lower-right panel of Figure~9), nevertheless the massive undersampling
casts many doubts on the true accuracy of derived radii. In fact, high
quality and massive ground-based observations of V505~Per by Marschall et
al. (1997) fully confirm GAIA-like solution of V505~Per as reported in
Figure~9 for all orbital and physical parameters but individual radii. If
the sum of them is R$_1$ + R$_2$ = 2.54 R$_\odot$ for the GAIA-like solution and
2.55 for Marschall et al., the individual values are 1.29 and 1.26 for the
latter and 1.40 and 1.14 R$_\odot$ for M01b. Numerical experiments however
show that by just doubling the number of points covering the principal
eclipses in the Hipparcos lightcurve would fix the result to a much higher
degree of confidence and close to the Marschall et al. findings.

It is therefore of great relevance to the broadest stellar astrophysics that
a minimum number of {\sl identical} photometric bands is replicated in all
the three viewing fields of GAIA, so to achieve maximum density of points in
lightcurve mapping of eclipsing binaries (with obvious benefit to all other types of
variable objects observed by GAIA).

The high quality of the results on V505~Per shown in Figure~9 (typical of
other cases investigated in M01b) is however definitively depending upon the
adopted 0.25\AA/pix spectral dispersion. At such a dispersion, lines from
both components are easily resolved and measured, while at the coarser 0.75
\AA/pix dispersion currently baselined for GAIA the lines would merge into
unresolved profiles that could even hide the binary nature of the object.

\subsection{Pulsating stars}

\begin{figure}[t]
\includegraphics[width=14.5cm]{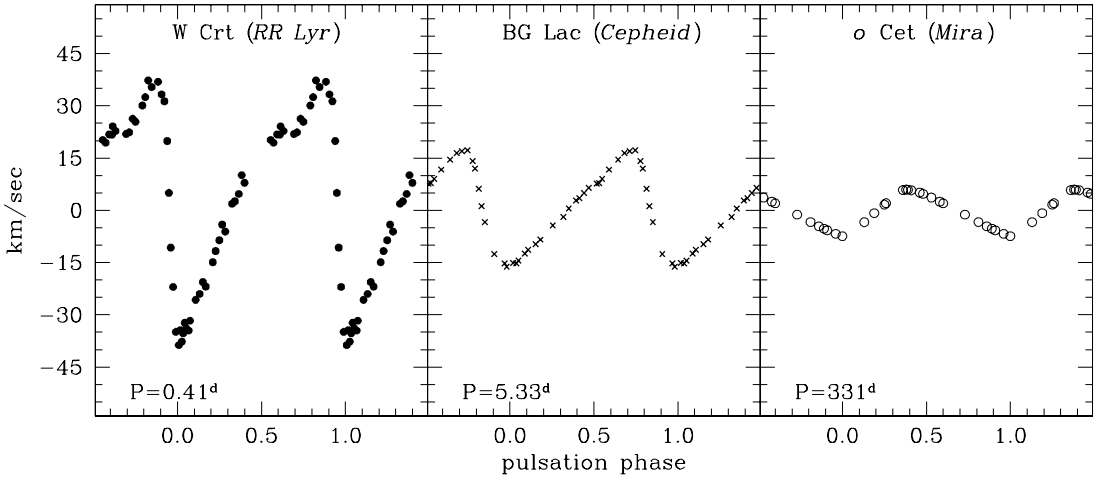}
\caption{Radial velocity curves of representative pulsating stars.}
\end{figure}

Radial velocity mapping of pulsating stars add precious information to the
effort of understanding their nature (cf. Bono et al. 1997). Non-radial
pulsators manifest marked variations of line profiles that can be
observed and modeled if the spectral resolution is high enough.

Spectra of template radially pulsating stars (RR~Lyr, $\delta$~Cep, $o$~Cet)
are presented in M99a. The wealth of strong absorption lines, in particular
Ca~II, assures that accurate radial velocities can be obtained at even the
lowest metallicities (RR~Lyr itself has [Z/Z$_\odot$]=$-$1.37).

Pulsation curves of representative cases are given in Figure~10. Shapes and
details of such curves (e.g. the glitch in W~Crt at phase 0.67, or BG~Lac at
phase 0.52), tell a lot about interior physics of these stars. It may be
easily anticipated that GAIA spectroscopic monitoring of thousands of them
(compared to the very few cases investigated from the ground) will loudly
impact on our understanding of stellar radial pulsations, provided the
accuracy of epoch radial velocities will be good enough.

Semi-amplitudes for Lyrids are of the order of $\bigtriangleup$RV$\sim$40
km~sec$^{-1}$, $\bigtriangleup$RV$\sim$15 for Cepheids and
$\bigtriangleup$RV$\sim$7 for Miras. So low amplitudes need high spectral
resolution to be properly mapped: at 0.25 \AA/pix even glitches in the
pulsation curves will be detected, while at 0.75 \AA/pix the pulsation curves
of the majority of Miras will be unobservable.

\begin{figure}[b]
\includegraphics[width=14.5cm]{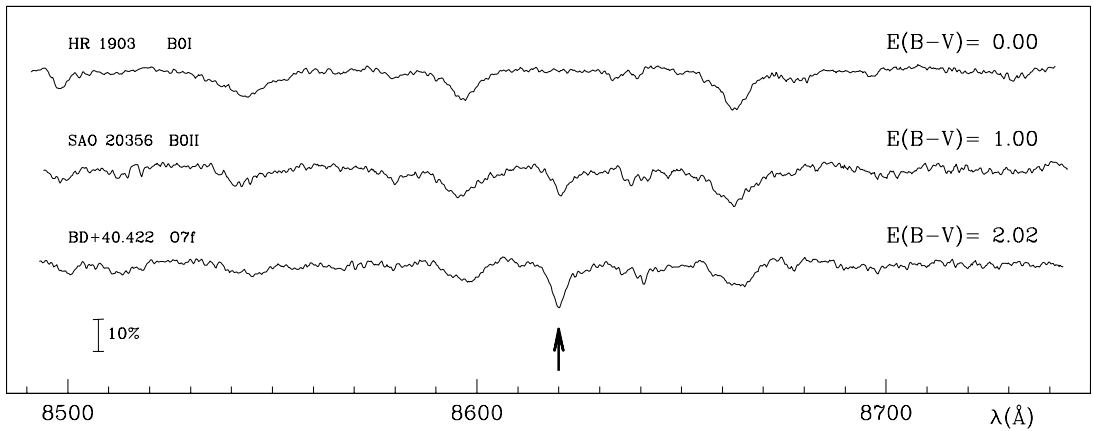}
\caption{The diffuse interstellar band at 8620 \AA\ (marked by the arrow) on
spectra of early type stars at different reddenings.
Note the increase of DIB equivalent width with reddening (from Munari
2000).}
\end{figure}

\section{Peculiarities}

The 8480$-$8750 \AA\ GAIA wavelength range offers fine detection
capabilities and diagnostic potential toward stellar peculiarities. Spectra
of a few representative peculiar stars are presented in Figure~12 (for other types
and examples see M99a) and briefly commented hereafter.

Be stars show strong and variable Paschen and He~I emission lines, which
profile trace conditions within the circumstellar disk. Hot mass-losing
stars (P-Cyg type) present Paschen and He~I lines with the characteristic
emission/absorption profile that can be modeled to derive the mass-loss
rate. Very wide and bright Ca~II profiles (and usually weaker Paschen) trace
the fast expansion of novae ejecta (FWHM=1280 km~sec$^{-1}$ in Nova Cyg 2001
\#1), with sub-structures correlated to and tracing the dishomogeneities and
clumps in the dispersing material.

Either hot and cool pre-ZAMS objects well perform over the GAIA range. Herbig Ae/Be
stars display both CaII and Paschen lines in strong emission, usually
emerging from an absorption core. In cooler pre-ZAMS objects Ca~II lines
play the game, from the exceptional intensity in T~Tau (cf. the spectrum in M99a)
to the weak emission components in FU~Ori.

Active atmosphere/spot stars reveal their nature by a complex cool
absorption spectrum and structured emission profiles. Major
bright and active areas over the stellar surface correlates with individual,
radial velocity displaced (by axial rotation) components in the spectrum, as
in the BY~Dra spectrum in Figure~12 where emission cores in Ca~II separated
by 74 km~sec$^{-1}$ are paralleled by Fe~I double absorption lines
separated by an identical 74 km~sec$^{-1}$.

\begin{figure}[Ht!]
\includegraphics[width=14.5cm]{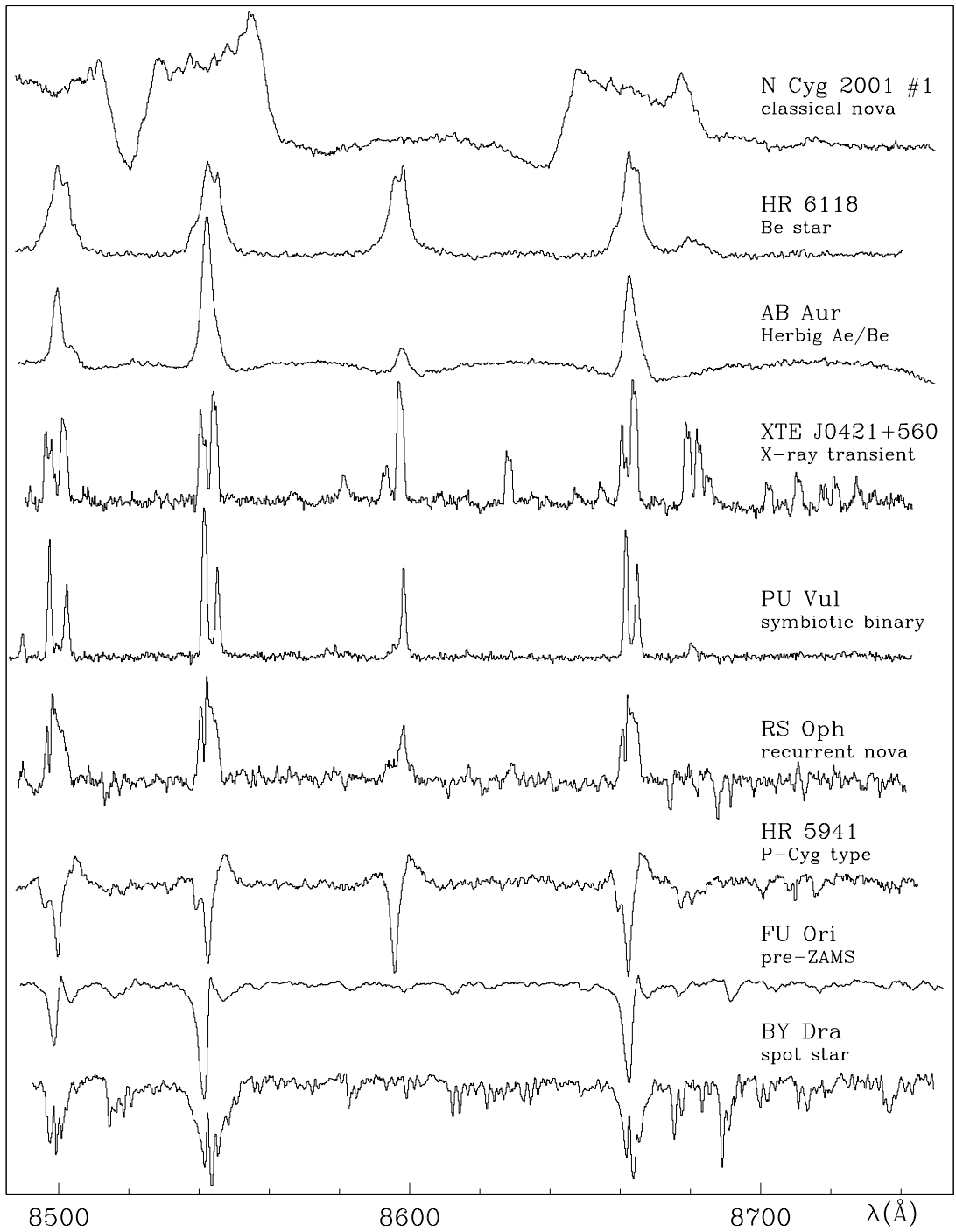}
\caption{Spectra of representative peculiar stars (from Munari et al. 2001c).}
\end{figure}

Spectra of interacting binary stars offer a fine and intriguing display over
the GAIA spectral range. Symbiotic stars usually present strong Paschen,
Ca~II and weaker He~I emission lines, while in some other case the naked
absorption spectrum of the cool giant can be observed without contamination
from the hot companion. The same Ca~II, Paschen and He~I lines plus strong
N~I (multiplet \#1 and \#8) dominate spectra of the X-ray transient XTE
J0421+560 in Figure~12, with complex line profiles indicating kinematically
decoupled emitting regions. Wide, strong and multi-component, orbital-phase
variable Ca~II profiles stand out in the spectra of the recurrent nova
RS~Oph.

Finally, the interstellar medium can manifest itself in GAIA spectra.
Figure~11 show the spectra of three similar hot stars affected by different
amount of reddening, where absorption by the diffuse interstellar band at 
8620 \AA\ clearly stands out. The equivalent width of the latter correlates
surprisingly well with reddening, offering interesting opportunities for
GAIA diagnostic of the interstellar medium (Munari 2000):
\begin{equation}
E_{B-V} \ = 2.69\ \times \ EW({\rm \AA})
\end{equation}
Efforts are currently underway (cf. Moro and Zwitter 1999) to investigate
the dependence of the slope coefficient (2.69) upon the
intrinsic and tri-dimensional properties of the galactic interstellar medium
($E_{B-V} \ = \ \alpha(l,b,{\rm D})\ \times \ EW({\rm \AA})$).

\end{document}